\documentclass[jfm,aps,preprint,showpacs,preprintnumbers,amsmath,amssymb,secnumroman,eqsecnum]{revtex4}

\usepackage{graphicx}
\usepackage{dcolumn}
\usepackage{bm}

\newcommand{\beq}{\begin{equation}}
\newcommand{\eeq}{\end{equation}}
\newcommand{\beqa}{\begin{eqnarray}}
\newcommand{\eeqa}{\end{eqnarray}}
\newcommand{\vc}[1]{\mbox{\boldmath $#1$}}

\newcommand{\vol}[1]{{\bf #1}}

\begin{document}


\title{Swimming at small Reynolds number of an elliptical disk propelled by an elliptically polarized surface wave}

\author{B. U. Felderhof}

 \email{ufelder@physik.rwth-aachen.de}
\affiliation{Institut f\"ur Theorie der Statistischen Physik\\ RWTH Aachen University\\
Templergraben 55\\52056 Aachen\\ Germany\\
}%

\date{\today}

\begin{abstract}
The swimming of an elliptical disk at small Reynolds number is studied on the basis of a perturbative solution of the Navier-Stokes equations for fluid flow near a deformable infinite sheet. A stroke involving an elliptically polarized plane surface wave is studied, in extension of work by Taylor and Tuck. In general the elliptic polarization of the stroke leads to an asymmetry of the flow in the upper and lower half-space. On the basis of results for an infinite sheet expressions for the mean translational and rotational swimming velocity of an elliptical disk of size much larger than the wavelength of the stroke are deduced. In addition expressions are derived for the mean power and the efficiency of swimming.
\end{abstract}

\maketitle
\section{\label{I}Introduction}

 In the study of swimming of a body immersed in a viscous incompressible fluid it may be assumed that the fluid flow is governed by the Navier-Stokes equations for flow velocity and pressure, and that a no-slip boundary condition holds at the body surface. The fluid is caused to move by periodic surface deformation of the body. Theory must explain that on average over a period the body acquires a steady swimming velocity. It also must provide expressions for the swimming velocity and the required power.

We assume in the following that the Reynolds number is small, so that the flow remains laminar. This requires small amplitude of deformation. The Reynolds number is defined as usual as $Re=UL/\nu$, where $U$ is the swimming velocity, $L$ is the body length, and $\nu$ is the kinematic viscosity of the fluid. To lowest order the swimming velocity is quadratic in the amplitude. A second independent parameter is the dimensionless viscosity. In the present context this is defined as $\zeta=k^2\nu/\omega$, where $k$ is the wavenumber and $\omega$ is the frequency of surface deformation.

The seminal work of Taylor \cite{1} was concerned with the swimming of a planar sheet with transverse plane wave surface deformation. He considered small amplitude deformation, corresponding to small Reynolds number, and the Stokes limit of high viscosity. Correspondingly the fluid equations of motion reduce to the Stokes equations. Taylor showed that to second order in the amplitude the swimming velocity is independent of viscosity. His work was generalized to the full range of dimensionless viscosity by Reynolds \cite{2} and by Tuck \cite{3}, who showed that then the swimming velocity depends on the dimensionless viscosity $\zeta$.

Taylor showed that in the Stokes limit the mean lateral force per unit area exerted by the pressure on the surface cancels against the mean force exerted by the viscous stress. This may be taken to imply that the mean thrust cancels against the mean drag. Conversely the force per unit area exerted by the sheet on the fluid vanishes. At smaller values of $\zeta$ the fluid is subject in addition to Reynolds stress, but the corresponding net mean force on the fluid per unit area again vanishes, as is seen by integration.

Tuck \cite{3} studied the effect of fluid inertia and considered also longitudinal plane wave squirming type surface deformations. In the Stokes limit the corresponding swimming velocity is in the opposite direction to that for transverse deformation. The swimming velocity depends again on the dimensionless viscosity $\zeta$, and the net force exerted on the fluid per unit area vanishes.

Blake \cite{4}, Brennen \cite{5}, and Childress \cite {6} studied flow generated by waving cilia with an envelope represented by a surface with deformations intermediate between transverse and longitudinal. We showed earlier in the Stokes limit that in the case of a sheet such intermediate surface deformation leads to an asymmetry between upper and lower half-space \cite{7}. This complicates the interpretation of the model and necessitates consideration of a body of finite size. We considered a circular disk and argued that the asymmetry corresponds to a mean rotational swimming velocity \cite{7}. The rotational velocity follows from the requirement that besides the net mean force also the net mean torque exerted on the fluid vanish.

In the following we consider arbitrary values of the dimensionless viscosity $\zeta$ and a planar sheet with deformation given by an elliptically polarized surface wave. There is an asymmetry between upper and lower half-space, except in the extreme limits considered by Tuck \cite{3}. Explicit expressions are derived for the mean translational and rotational swimming velocity and the power for an elliptical disk of size much larger than the wavelength of deformation. The results may be compared with those derived for a slab with symmetric surface deformation leading to symmetry of the flow on both sides \cite{8}.

\section{\label{2}Swimming elliptical disk}

We consider an elliptical disk immersed in a viscous incompressible fluid of shear viscosity $\eta$ and mass density $\rho$. The rest shape of the disk is denoted as $S_0$. The fluid is set in motion by time-dependent distortions of the disk. We assume that the distortions are periodic in time with period $T=2\pi/\omega$. We consider a prescribed time-dependent shape $S(t)$ leading to swimming motion. The flow velocity
$\vc{v}(\vc{r},t)$ and the pressure $p(\vc{r},t)$ in the rest frame of the disk satisfy the
Navier-Stokes equations with an added Coriolis force term \cite{9} corresponding to the mean rotational swimming velocity $\vc{\Omega}$.

The surface displacement
$\vc{\xi}(\vc{s},t)$ is defined as the vector distance
\begin{equation}
\label{2.1}\vc{\xi}=\vc{s}'-\vc{s}
\end{equation}
of a point $\vc{s}'$ on the displaced surface $S(t)$ from the
point $\vc{s}$ on the disk $S_0$. The fluid
velocity $\vc{v}(\vc{r},t)$ is required to satisfy the no-slip boundary condition
\begin{equation}
\label{2.2}\vc{v}(\vc{s}+\vc{\xi}(\vc{s},t))=\frac{\partial\vc{\xi}(\vc{s},t)}{\partial t}.
\end{equation}

We construct an approximate perturbative solution to the Navier-Stokes equations with boundary condition (2.2) by formal expansion \cite{9} of the flow velocity and the pressure in powers of $\vc{\xi}$,
\begin{equation}
\label{2.3}\vc{v}=\vc{v}_1+\vc{v}_2+...,\qquad p=p_0+p_1+p_2+....
\end{equation}
Correspondingly the mean translational and rotational swimming velocities are expanded as
\begin{equation}
\label{2.4}\vc{U}=\vc{U}_1+\vc{U}_2+....,\qquad\vc{\Omega}=\vc{\Omega}_1+\vc{\Omega}_2+.....
\end{equation}
The flow velocity at the surface is formally expanded as
\begin{equation}
\label{2.5}\vc{v}(\vc{s}+\vc{\xi}(\vc{s},t))=\vc{v}(\vc{s},t)+(\vc{\xi}\cdot\nabla)\vc{v}(\vc{r},t)\big|_{\vc{r}=\vc{s}}+....
\end{equation}
With the aid of this expansion the boundary condition may be applied at the undisplaced surface.

To first order the fluid equations of motion reduce to the linearized Navier-Stokes equations
\begin{equation}
\label{2.6}\rho\frac{\partial\vc{v}_1}{\partial t}=\eta\nabla^2\vc{v}_1-\nabla p_1,\qquad\nabla\cdot\vc{v}_1=0.
\end{equation}
The first order boundary condition is
\begin{equation}
\label{2.7}\vc{v}_1(\vc{s},t)=\frac{\partial\vc{\xi}(\vc{s},t)}{\partial t},\qquad\vc{s}\in S_0.
\end{equation}
The first order mean swimming velocities $\vc{U}_1$ and $\vc{\Omega}_1$ vanish.

To second order the time-averaged flow $\overline{\vc{v}_2},\overline{p_2}$ satisfies the inhomogeneous Stokes equations  \cite{9}
\begin{equation}
\label{2.8}\eta\nabla^2\overline{\vc{v}}_2-\nabla\overline{p_2}=-\overline{\vc{F}_2},\qquad\nabla\cdot\overline{\vc{v}_2}=0,
\end{equation}
with force density
\begin{equation}
\label{2.9}\overline{\vc{F}_2}=-\rho\overline{(\vc{v}_1\cdot\nabla)\vc{v}_1},
\end{equation}
where the overline indicates averaging over a period $T=2\pi/\omega$. The force density may be written as the divergence of a Reynolds stress tensor. The Stokes equations Eq. (2.8) must be solved with the boundary condition
\begin{equation}
\label{2.10}\overline{\vc{v}}_2(\vc{s})=-\overline{(\vc{\xi}\cdot\nabla)\vc{v}_1}\;\big|_{\vc{s}},
\end{equation}
 as follows from Eqs. (2.2) and (2.5).

We use Cartesian coordinates $x,y,z$ such that the disk at rest is in the plane $y=0$ and centered at the origin. In our calculation the disk is approximated by an infinite sheet with displacements which do not depend on $z$ and take the form $\vc{\xi}=(\xi_{ x}(x,t),\xi_{y}(x,t),0)$ for $x$ axis either in the direction of the long or short axis of the disk. As a consequence the flow velocity $\vc{v}$ and pressure $p$ do not depend on $z$, and the problem is effectively two-dimensional in the $x,y$ coordinates. We shall consider situations where the mean flow velocity $\overline{\vc{v}_2}(\vc{r})=\overline{v_2}\vc{e}_x$ depends only on $y$, and tends to a constant $-U_{2\pm}\vc{e}_x$ as $y$ tends to $\pm\infty$. At a later stage we shall associate the two prefactors $U_{2\pm}$ with the second order mean translational and rotational swimming velocity of the disk. The force density $\overline{\vc{F}_2}$ is of the form $\overline{\vc{F}_2}=\overline{F_{2x}}(y)\vc{e}_x+\overline{F_{2y}}(y)\vc{e}_y$ and the mean pressure $\overline{p_2}$ depends only on $y$.

We consider an undulating sheet with plane wave distortion along the $x$ direction. It will be convenient to use complex notation. Thus we write for the displacement vector
\begin{equation}
\label{2.11}\vc{\xi}(x,t)=\vc{\xi}^c e^{ikx-i\omega t},
\end{equation}
with complex amplitude $\vc{\xi}^c$ and with the understanding that the real part of the expression is used to get the physical displacement. Correspondingly the first order velocity $\vc{v}_1(\vc{r},t)$ and pressure $p_1(\vc{r},t)$ in the upper and lower half-space take the form
\begin{equation}
\label{2.12}\vc{v}_1(\vc{r},t)=\vc{v}^c_\pm(y) e^{ikx-i\omega t},\qquad p_1(\vc{r},t)=p^c_\pm(y) e^{ikx-i\omega t}.
\end{equation}
The boundary value in Eq. (2.10) may be expressed as
\begin{equation}
\label{2.13}\overline{\vc{v}}_2(\vc{s})=-\frac{1}{2}\mathrm{Re}\;[(\vc{\xi}^{*}\cdot\nabla)\vc{v}_1\big|_{y=\pm 0}],
\end{equation}
with values taken on either the upper or lower side of the sheet.

The power required equals the rate of dissipation of energy in the fluid. This can be calculated from the work done per unit area against the stress $\vc{\sigma}=\eta(\nabla\vc{v}+(\nabla\vc{v})^T)-p\vc{I}$. The mean rate of dissipation per unit area is to second order \cite{9}
\begin{equation}
\label{2.14}\overline{D}_2=-\frac{1}{2}\mathrm{Re}\;[\vc{v}^*_{1+}\cdot\vc{\sigma}_{1+}\cdot\vc{e}_y\big|_{y=0+}]
+\frac{1}{2}\mathrm{Re}\;[\vc{v}^*_{1-}\cdot\vc{\sigma}_{1-}\cdot\vc{e}_y\big|_{y=0-}].
\end{equation}

We define the dimensionless efficiency as
\begin{equation}
\label{2.15}E_2=4\eta\omega\frac{|U_2|}{\overline{D}_2}.
\end{equation}
We use the prefactor $4$ so that this equals unity in the resistive limit for the situation considered by Taylor \cite{1}.

\section{\label{III}Reynolds force density}

Tuck \cite{3} generalized Taylor's solution for an undulating thin sheet to the case of a fluid with inertia. Elsewhere \cite{8} we derived his results from Eqs. (2.8)-(2.10). In Taylor's solution the displacement vector is given by Eq. (2.11) with $\vc{\xi}^c=A\vc{e}_y$ with real amplitude $A$. The first order flow velocity and pressure are given by Eq. (2.12) with
\begin{eqnarray}
\label{3.1}v^c_{x\pm}(y)&=&\mp B\omega\big(e^{\mp ky}-e^{\mp sy}\big) ,\qquad v^c_{y\pm}(y)=-iB\omega\big(e^{\mp ky}-\frac{k}{s}\;e^{\mp sy}\big),\nonumber\\
p^c_{\pm}(y)&=&\mp\frac{\omega^2\rho}{k}Be^{\mp ky},\qquad s=\sqrt{k^2-\frac{i\omega\rho}{\eta}},\qquad B=\frac{s}{s-k}A,
\end{eqnarray}
where the upper (lower) sign refers to the upper (lower) half-space.

Tuck considered also a situation with squirming displacements given by Eq. (2.11) with $\vc{\xi}^c=A\vc{e}_x$ with real amplitude $A$. In this case the first order flow velocity and pressure are given by Eq. (2.12) with
\begin{eqnarray}
\label{3.2}v^c_{x\pm}(y)&=&-iB\omega\big(e^{\mp ky}-\frac{s}{k}\;e^{\mp sy}\big) ,\qquad v^c_{y\pm}(y)=\pm B\omega\big(e^{\mp ky}-e^{\mp sy}\big),\nonumber\\
p^c_{\pm}(y)&=&-i\frac{\omega^2\rho}{k}Be^{\mp ky},\qquad B=\frac{k}{k-s}A.
\end{eqnarray}

We consider more generally a situation where the surface displacement is given by Eq. (2.11) with
complex amplitude vector
\begin{equation}
\label{3.3}\vc{\xi}^c=A\cos\alpha\;\vc{e}_x+iA\sin\alpha\;\vc{e}_y,
\end{equation}
with real amplitude $A$ with dimension length, corresponding to an elliptically polarized plane surface wave. Tuck's generalization of Taylor's solution corresponds to $\alpha=\pi/2$, and his squirming wave corresponds to $\alpha=0$.

It turns out that for intermediate values of $\alpha$ the mean second order velocity component $\overline{v_{2x}}(y)$ and the mean second order pressure $\overline{p_2}(y)$ are not symmetric in $y$. Only in the cases $\alpha=0$ and $\alpha=\pi/2$ are these quantities symmetric. We shall argue that the asymmetry corresponds to a rotational swimming velocity of a flat elliptical disk which is approximated by the infinite sheet.

The mean swimming velocity depends on the fluid properties $\eta$ and $\rho$ only via the dimensionless viscosity
\begin{equation}
\label{3.4}\zeta=\frac{\eta k^2}{\omega\rho}.
\end{equation}
The Stokes limit corresponds to large values of $\zeta$, and the inertia-dominated regime corresponds to small values. It is of interest to explore the properties of the model in the full range of $\zeta$. Tuck \cite{3} and Childress \cite{6} call $1/\zeta$ the Reynolds number $R$.

In complex notation the mean Reynolds force density $\overline{\vc{F}_2}$ in Eq. (2.9) can be expressed as
\begin{equation}
\label{3.5}\overline{\vc{F}_2}=-\frac{1}{2}\mathrm{Re}\;[\rho(\vc{v}^*_1\cdot\nabla)\vc{v}_1].
\end{equation}
The $x$-component has the value
\begin{equation}
\label{3.6}\overline{F_{2x}}(y)=\frac{1}{2}A^2\rho\omega^2\mathrm{Re}\;\big[(1\mp\sin2\alpha)f_{2x1\pm}(y)+\cos2\alpha\; f_{2x2\pm}(y)\big],
\end{equation}
with functions
\begin{eqnarray}
\label{3.7}f_{2x1\pm}(y)&=&\frac{i}{2|k-s|^2}\bigg[k|k+s|^2 e^{\mp(k+s^*)y}+s|k+s|^2e^{\mp(k+s)y}-2ks(s+s^*)e^{\mp(s+s^*)y}\bigg],\nonumber\\
f_{2x2\pm}(y)&=&\frac{i}{2|k-s|^2}\bigg[k(k-s)(k+s^*)e^{\mp(k+s^*)y}+s(k+s)(k-s^*)e^{\mp(k+s)y}\bigg].
\end{eqnarray}
At $y=0$ the functions take the values,
\begin{equation}
\label{3.8}f_{2x1+}(0+)=f_{2x1-}(0-),\qquad f_{2x2+}(0+)=f_{2x2-}(0-).
\end{equation}
The force density component $\overline{F_{2x}}(y)$ is discontinuous at $y=0$ because of the factor $\;\;\;\;\;\;\;\;\;\;\;\;(1\mp\sin2\alpha)$ in Eq. (3.6).

The corresponding solution of Eq. (2.8) takes the form $\overline{\vc{v}_2}=\overline{v_{2x}}(y)\vc{e}_x$ with $x$-component
\begin{equation}
\label{3.9}\overline{v_{2x}(y)}=-U_{2\pm}
-\frac{1}{2\eta}A^2\rho\omega^2\mathrm{Re}\;\big[(1\mp\sin2\alpha)u_{2x1\pm}(y)+\cos2\alpha\; u_{2x2\pm}(y)\big],
\end{equation}
with functions
\begin{eqnarray}
\label{3.10}u_{2x1\pm}(y)&=&\mp\frac{i}{2|k-s|^2}\bigg[k\frac{k+s}{k+s^*} e^{\mp(k+s^*)y}+s\frac{k+s^*}{k+s}e^{\mp(k+s)y}-2\frac{ks}{s+s^*}e^{\mp(s+s^*)y}\bigg],\nonumber\\
u_{2x2\pm}(y)&=&\mp\frac{i}{2|k-s|^2}\bigg[k\frac{k-s}{k+s^*}e^{\mp(k+s^*)y}+s\frac{k-s^*}{k+s}e^{\mp(k+s)y}\bigg].
\end{eqnarray}
From Eq. (2.13) we find the boundary condition
\begin{equation}
\label{3.11}\overline{v_{2x}}(0\pm)=-\frac{1}{4}A^2\omega\;\mathrm{Re}\;\big[k-s\pm(k+s)\sin2\alpha+(k+s)\cos2\alpha\big].
\end{equation}
Substituting this into Eq. (3.9) we find the velocities $U_{2\pm}$. The boundary values in Eq. (3.11) demonstrate the asymmetry mentioned above. In Fig. 1 we show the dimensionless mean flow velocity $\overline{v_{2x}}(y)/(A^2\omega k)$ at $\alpha=\pi/4$ and $\zeta=1$. At this value of $\alpha$, corresponding to a circularly polarized surface wave, $\sin2\alpha=1$ and $\cos2\alpha=0$, so that the second term in Eq. (3.9) vanishes for $y>0$ and the velocity reduces to $-U_{2+}$.

The force density component $\overline{F_{2x}}(y)$ is asymmetric in $y$. For each of the functions in Eq. (3.7) we find
\begin{eqnarray}
\label{3.12}\mathrm{Re}\int^\infty_0f_{2x1+}(y)\;dy&=&0,\qquad\mathrm{Re}\int^\infty_0f_{2x2+}(y)\;dy=0,\nonumber\\
\mathrm{Re}\int_{-\infty}^0f_{2x1-}(y)\;dy&=&0,\qquad\mathrm{Re}\int_{-\infty}^0f_{2x2-}(y)\;dy=0,
\end{eqnarray}
so that
\begin{equation}
\label{3.13}\int^\infty_0\overline{F_{2x}}(y)\;dy=0,\qquad\int^0_{-\infty}\overline{F_{2x}}(y)\;dy=0.
\end{equation}
For the total moment we find
\begin{eqnarray}
\label{3.14}Y_F&=&\int^\infty_{-\infty} y\overline{F_{2x}}(y)\;dy\nonumber\\
&=&\frac{1}{4}A^2\rho\omega^2\sin2\alpha\;\frac{i(2k+s+s^*)(s-s^*)}{|k+s|^2(s+s^*)}.
\end{eqnarray}
In particular this shows that $Y_F$ vanishes in the symmetric cases $\alpha=0$ and $\alpha=\pi/2$. Furthermore $Y_F$ vanishes for any $\alpha$ in the Stokes limit, since then $s=s^*=k$. By use of $s=k\sqrt{1-i/\zeta}$ we can rewrite Eq. (3.14) as
\begin{equation}
\label{3.15}Y_F=A^2\frac{\rho\omega^2}{2k}\;g(\zeta)\sin2\alpha,
\end{equation}
with the dimensionless function
\begin{equation}
\label{3.16}g(\zeta)=\frac{\sqrt{F^2-1}}{2F^2},
\end{equation}
in terms of Tuck's parameter $F$ defined by
\begin{equation}
\label{3.17}F=\frac{1}{\sqrt{2}}\sqrt{1+\sqrt{1+\frac{1}{\zeta^2}}},\qquad \zeta=\frac{1}{\sqrt{4F^4-4F^2}}.
\end{equation}
In Fig. 2 we plot the function $g(\zeta)$ as a function of $\xi=-\log_{10}\zeta$.
The function $g(\zeta)$ takes its maximum value $1/4$ at $\zeta_m=1/(2\sqrt{2})=0.354$.

We consider also the separate moments
\begin{equation}
\label{3.18}Y_{F+}=\int^\infty_0 y\overline{F_{2x}}(y)\;dy,\qquad Y_{F-}=\int_{-\infty}^0 y\overline{F_{2x}}(y)\;dy,
\end{equation}
with sum $Y_F=Y_{F+}+Y_{F-}$. We write in analogy to Eq. (3.15)
\begin{equation}
\label{3.19}Y_{F\pm}=A^2\frac{\rho\omega^2}{4k}\bigg[g(\zeta)(\mp 1+\sin2\alpha)\pm h(\zeta)\cos2\alpha\bigg],
\end{equation}
with function $h(\zeta)$ given by
\begin{equation}
\label{3.20}h(\zeta)=\frac{F-1}{2F\sqrt{F^2-1}}.
\end{equation}
In Fig. 3 we plot the function $h(\zeta)$ as a function of $\xi=-\log_{10}\zeta$. The maximum value is $0.150$ at $\zeta_m=1/(2\sqrt{2+\sqrt{5}})=0.243$.

By use of the Stokes equation $\eta d^2\overline{v_{2x}}/dy^2=-\overline{F_{2x}}$ and an integration by parts in Eq. (3.18) we find the relations
\begin{equation}
\label{3.21}Y_{F+}=\eta[-U_{2+}-\overline{v_{2x}}(0+)],\qquad Y_{F-}=\eta[U_{2-}+\overline{v_{2x}}(0-)].
\end{equation}
These relations will be useful in the next section.

One can also evaluate the $y$- component of the force density $\overline{\vc{F}_2}$. This is balanced by the mean second order pressure $\overline{p}_2(y)$ in the Stokes equation
\begin{equation}
\label{3.22}\frac{d\overline{p}_2}{dy}=\overline{F_{2y}}.
\end{equation}
Integrating over $y$ we find
\begin{equation}
\label{3.23}\int^\infty_{-\infty}\overline{F_{2y}}(y)\;dy=0.
\end{equation}
The contributions from the upper and lower half-space cancel. They are given by the simple expressions
\begin{equation}
\label{3.24}\int^\infty_0\overline{F_{2y}}(y)\;dy=-\int_{-\infty}^0\overline{F_{2y}}(y)\;dy=\frac{1}{2}A^2\rho\omega^2\sin^2\alpha,
\end{equation}
independent of viscosity.

It follows from Eq. (3.23) that we can identify
\begin{equation}
\label{3.25}n_{zF}=-\int^\infty_{-\infty}y\overline{F_{2x}}(y)\;dy=-Y_F
\end{equation}
as the $z$-component of the torque exerted on the fluid by the Reynolds force density $\overline{\vc{F}_2}$, per unit area of the sheet. It follows from Eq. (3.15) that this vanishes only in the Stokes limit $\zeta\rightarrow\infty$, in the inertia-dominated limit $\zeta\rightarrow 0$, and in the symmetric cases $\alpha=0$ and $\alpha=\pi/2$.

From Eq. (2.14) we find
\begin{equation}
\label{3.26}\overline{D_2}=A^2\eta\omega^2k(1+F).
\end{equation}
This is independent of $\alpha$ and agrees with Tuck's expressions \cite{3} for $\alpha=0$ and $\alpha=\pi/2$. The contributions from the two half-spaces differ, except for $\alpha=0$ and $\alpha=\pi/2$. In Fig. 4 we plot the efficiency $E_2$, defined in Eq. (2.15), as a function of $\alpha$ and $\xi=-\log_{10}\zeta$. In the Stokes limit this tends to $|\cos2\alpha|$.

It is of interest to calculate also the mean kinetic energy of flow. We find to second order
\begin{equation}
\label{3.27}\overline{K_2}=\frac{1}{4}\;\rho\int^\infty_{-\infty}|\vc{v}_1|^2\;dy=A^2\frac{\rho\omega^2}{4k}\frac{1+F-F\cos2\alpha}{F}.
\end{equation}
For the symmetric cases $\alpha=0$ and $\alpha=\pi/2$ this agrees with the expressions found earlier \cite{8}. For other values of $\alpha$ the contributions from the upper and lower half-space differ.

Finally, we mention one further property of the mean second order flow pattern. It follows from Eq. (3.22) that the mean second order pressure profile can be calculated from the mean Reynolds force density component $\overline{F_{2y}}(y)$. In Fig. 5 we show as an example the reduced profile $\hat{p}_2(y)=\overline{p_2}(y)/(A^2\rho\omega^2)$ for $\alpha=\pi/4$ and $\zeta=1$. As expected, this is again asymmetric.

\section{\label{IV}Translational and rotational swimming of a disk}

In this section we discuss the translational and rotational swimming of an elliptical disk with semi-axes $a$ and $b$, based on the properties of the solution for a sheet found in the preceding two sections. It is assumed that $kb>>1$, so that the surface deformations of the disk have a wavelength $\lambda=2\pi/k$ much smaller than the size of the disk. In the cylindrical volume consisting of the disk and a few wavelengths on both sides the flow pattern is well approximated by that of an infinite sheet. This implies that at a distance of a few wavelengths the mean flow velocity tends to the constant $\overline{\vc{v}_2}(y)=-U_{2\pm}\vc{e}_x$ in the rest frame of the disk. The mean second order swimming velocity can be identified as
\begin{equation}
\label{4.1}U_2=\frac{1}{2}\;(U_{2+}+U_{2-}).
\end{equation}
From Eqs. (3.9) and (3.11) we find
\begin{equation}
\label{4.2}U_2=\frac{1}{4}A^2\omega k\frac{-1+F+2F\cos2\alpha}{F}.
\end{equation}
The result agrees with Tuck's expressions in the symmetric cases $\alpha=0$ and $\alpha=\pi/2$. In Fig. 6 we plot the reduced swimming velocity $\hat{U}_2=2U_2/(A^2\omega k)$ as a function of $\alpha$ and $\xi=-\log_{10}\zeta$. We have normalized such that $\hat{U}_2=-1$ for $\alpha=\pi/2$ and $\zeta\rightarrow\infty$, corresponding to Taylor's solution \cite{1}. The maximum value $\hat{U}_2=3/2$ is reached for $\alpha=0$ and $\zeta\rightarrow 0$. It is seen from Fig. 4 that in this limit the efficiency $E_2$ tends to zero, which suggests that in the low viscosity regime a different definition of the efficiency of stroke would be more appropriate. In Fig. 6 we also plot the curve on which $\hat{U}_2$ vanishes. In Fig. 7 we show $-\hat{U}_2$ for $\alpha=\pi/2$, and $\hat{U}_2$ for $\alpha=0$ and $\alpha=\pi/4$.

In Eq. (3.11) we found that in general the velocity component $\overline{v_{2x}}(y)$ is discontinuous at $y=0$, with symmetry only for $\alpha=0$ and $\alpha=\pi/2$. The discontinuity corresponds to a force dipole density. In addition to the bulk force density with component $\overline{F_{2x}}(y)$, given by Eq. (3.6), the mean force density component has a surface term $\overline{F_{S2x}}(y)=G\delta'(y)$ with amplitude $G$. From the Stokes equations Eq. (2.8) and Eq. (3.11) we find \cite{7}
\begin{equation}
\label{4.3}G=\eta[-\overline{v_{2x}}(0+)+\overline{v_{2x}}(0-)]=\frac{1}{2}A^2\eta\omega\sin2\alpha\;\mathrm{Re}[k+s].
\end{equation}
For an elliptical disk the force dipole density corresponds to a torque
\begin{equation}
\label{4.4}\vc{N}_S=\pi abG\vc{e}_z.
\end{equation}
This must be added to the torque exerted by the bulk force density, as given by Eq. (3.25). This yields the total torque
\begin{equation}
\label{4.5}\vc{N}=\pi ab (n_{zF}+G)\vc{e}_z.
\end{equation}
By use of Eqs. (3.21), (3.25), and (4.3) we find
\begin{equation}
\label{4.6}n_{zF}+G=\eta[U_{2+}-U_{2-}].
\end{equation}

At large distances from the disk the flow generated by the total mean force density becomes identical with that of a point torque,
\begin{equation}
\label{4.7}\hat{\vc{v}}_2(\vc{r})\approx\frac{1}{8\pi\eta}\vc{N}\times\frac{\vc{r}}{r^3},
\qquad\hat{p}_2(\vc{r})\approx 0,\qquad\mathrm{as}\;r\rightarrow\infty.
\end{equation}
In order to find the complete second order flow pattern at large distances we must add the Stokes flow which vanishes at the surface of the disk \cite{9}, and which cancels the term shown in Eq. (4.7). This flow has the asymptotic behavior
  \begin{equation}
\label{4.8}\vc{v}^{\mathrm{St}}_2(\vc{r})\approx -\vc{\Omega}_2\times\vc{r}-\frac{1}{8\pi\eta}\vc{N}\times\frac{\vc{r}}{r^3},
\qquad p^{\mathrm{St}}_2(\vc{r})\approx 0,\qquad\mathrm{as}\;r\rightarrow\infty,
\end{equation}
where $\vc{\Omega}_2$ is related to $\vc{N}$ by
  \begin{equation}
\label{4.9}\vc{N}=-f_{Ra,b}\vc{\Omega}_2,
\end{equation}
where $f_{Ra,b}$ is the rotational friction coefficient of an elliptical disk, with $f_{Ra}$ for $\vc{\Omega}_2$ in the direction of the long axis of the ellipse, or $f_{Rb}$ for $\vc{\Omega}_2$ in the direction of the short axis.

The rotational friction coefficients of the disk can be obtained from those for an ellipsoid \cite{10},\cite{11} by taking the limit $c\rightarrow 0$, where $c$ is the shortest axis of the ellipsoid. This yields
  \begin{equation}
\label{4.10}f_{Ra}=\frac{16\pi\eta a^3}{3B_0(q)},\qquad f_{Rb}=\frac{16\pi\eta a^3}{3A_0(q)},
\end{equation}
with dimensionless factors
  \begin{eqnarray}
\label{4.11}A_0(q)&=&\frac{2}{1-q}\big[K(1-q)-E(1-q)\big],\nonumber\\
B_0(q)&=&\frac{2}{1-q}\big[\frac{1}{q}\;E(1-q)-K(1-q)\big],\qquad q=\frac{b^2}{a^2},
\end{eqnarray}
with complete elliptic integrals $K(m)$ and $E(m)$ of the first and second kind \cite{12}. The factors have the properties $A_0(1)=B_0(1)=\pi/2$ corresponding to the rotational friction coefficient $f_R=32\eta a^3/3$ of a circular disk \cite{13}. In Fig. 8 we plot the ratios $f_{Ra}/f_R$ and $f_{Rb}/f_R$ a functions of $b/a$.

The complete mean second order flow tends to
 \begin{equation}
\label{4.12}\overline{\vc{v}}_2(\vc{r})\approx-\vc{\Omega}_2\times\vc{r}-U_2\vc{e}_x,
\qquad\overline{p}_2(\vc{r})\approx 0,\qquad\mathrm{as}\;r\rightarrow\infty.
\end{equation}
On both sides near the disk the mean flow pattern is nearly that of the infinite sheet, since the added Stokes solution satisfies the no-slip boundary condition.
From Eq. (4.12) we identify $\vc{\Omega}_2$ as the mean rotational swimming velocity of the disk in the laboratory frame. For surface wave in the direction of the long axis of the disk the vector $\vc{\Omega}_2$ is in the direction of the short axis, and for surface wave in the direction of the short axis of the disk the vector $\vc{\Omega}_2$ is in the direction of the long axis. From Eqs. (4.5), (4.6), and (4.9) we find
  \begin{equation}
\label{4.13}\vc{\Omega}_2=-\frac{\pi ab\eta}{f_{Ra,b}}\;(U_{2+}-U_{2-})\vc{e}_z.
\end{equation}
This shows that on time average the disk rotates steadily about a diameter perpendicular to the direction of translational swimming. Since the translational swimming velocity is always parallel to the plane of the disk this implies that on  a slow time scale the center of the disk runs through a circular orbit in the laboratory frame \cite{7}. The circle has radius $|U_2|/|\Omega_2|$ and is traversed in time $2\pi/|\Omega_2|$. At any point the disk is orthogonal and tangential to the circle.

The reduced mean rotational swimming velocity $\hat{\Omega}_2$ is given by
\begin{equation}
\label{4.14}\hat{\Omega}_2=\frac{U_{2+}-U_{2-}}{A^2\omega k\sin2\alpha}=\frac{1+F}{2F}.
\end{equation}
In Fig. 8 we plot $\hat{\Omega}_2$ as a function of $\xi=-\log_{10}\zeta$. For large $\zeta$, corresponding to the Stokes limit, $\hat{\Omega}_2$ varies as
\begin{equation}
\label{4.15}\hat{\Omega}_2=1-\frac{1}{16\zeta^2}+O(\zeta^{-4}).
\end{equation}
In the inertia-dominated regime it varies as
\begin{equation}
\label{4.16}\hat{\Omega}_2=\frac{1}{2}+\frac{1}{2}\sqrt{2\zeta}+O(\zeta^{3/2}).
\end{equation}
The mean rotational swimming velocity increases monotonically as a function of dimensionless viscosity $\zeta$.

\section{\label{VII}Discussion}

The analysis shows that an elliptical disk with elliptically polarized surface deformation has interesting swimming properties. We have found that for a general elliptically polarized stroke the disk acquires both a translational and a rotational mean swimming velocity. This implies that the disk can manoeuver by varying its stroke. In particular, the straight swimming suggested by the calculations of Taylor \cite{1} and Tuck \cite{3} for an infinite sheet with transverse plane wave surface deformation can be given a turn by making the stroke slightly elliptic.

It is noteworthy that the expression Eq. (4.2) for the mean translational swimming velocity of the disk does not involve its translational friction coefficient. The swimming velocity can always be written as the product of the friction coefficient and another factor \cite{15}, but this does not always make sense \cite{16}.

The requirements of small Reynolds number and a large value $kb$ still leave an interesting range of application. If we write the amplitude $A$ in Eq. (3.3) as $A=\varepsilon\lambda$, where $\varepsilon$ is a dimensionless factor and $\lambda=2\pi/k$ is the wavelength of deformation, then the translational swimming velocity is approximately $2\pi^2\varepsilon^2\omega/k$, corresponding to Reynolds number $Re\approx 4\pi^2\varepsilon^2ka/\zeta$. With $\varepsilon=0.1,\;ka=100$ and moderate Reynolds number $Re<1000$ for which the flow may be expected to be laminar, this still leaves a wide range of dimensionless viscosity $\zeta=\eta k^2/(\omega\rho)$ for which the theory may be applied. It has been suggested \cite{14} that in aquatic animal swimming the boundary layer is laminar even at high Reynolds number.

Experimentally the elliptically polarized deformation wave cannot be realized easily. In a recent experiment \cite{17} on swimming of a rectangular undulating plate the surface deformation is generated by a line source and determined by the elasticity of the plate. In the experiment $Re\approx 10^4$, and there is an appreciable wake, so that the above theory does not apply. The authors find agreement with the theory for swimming of an elongated object at large Reynolds number developed by Lighthill \cite{18},\cite{19}. Other interesting experiments involve horizontal rotational motion of a rigid or flexible rectangular plate oscillating vertically \cite{20},\cite{21}. The rotational motion is referred to as "forward flight".

It should be possible to perform computer simulations in the regime where the present theory applies. It would be of interest to compare with theory for the infinite sheet in simulations with periodic boundary conditions \cite{22},\cite{23}.  A review of recent numerical work by different methods was provided by Deng et al. \cite{24}.

The computer simulations can be performed in two dimensions, with the infinite sheet reduced to a line and the disk reduced to a line element, corresponding to a flexible wing of finite chord and infinite span. Since the flow is limited to a small range in the transverse $y$ direction, it suffices to use a rather small domain in this direction. It would be of interest to explore stability of the flow as a function of amplitude $\varepsilon$ and dimensionless viscosity $\zeta$. Computer simulations for a heaving and pitching rigid ellipsoidal foil of infinite span show the important effect of vortex shedding on swimming at large amplitude of stroke \cite{25}-\cite{27}. Such effects are beyond the low order perturbation theory studied in this article.

\newpage

\section*{Figure captions}

\subsection*{Fig. 1}
Plot of the dimensionless mean flow velocity $\overline{v_{2x}}(y)/(A^2\omega k)$ given by Eq. (3.9) at $\alpha=\pi/4$ and $\zeta=1$.

\subsection*{Fig. 2}
Plot of the dimensionless function $g(\zeta)$ given by Eq. (3.16) as a function of $\xi=-\log_{10}\zeta$.

\subsection*{Fig. 3}
Plot of the dimensionless function $h(\zeta)$ given by Eq. (3.20) as a function of $\xi=-\log_{10}\zeta$.

\subsection*{Fig. 4}
Plot of the efficiency $E_2$, defined in Eq. (2.15), as a function of polarization angle $\alpha$ and $\xi=-\log_{10}\zeta$.

\subsection*{Fig. 5}
Plot of the reduced profile $\hat{p}_2(y)=\overline{p_2}(y)/(A^2\rho\omega^2)$ for $\alpha=\pi/4$ and $\zeta=1$.

\subsection*{Fig. 6}
Plot of the reduced swimming velocity $\hat{U}_2=2U_2/(A^2\omega k)$ as a function of $\alpha$ and $\xi=-\log_{10}\zeta$. On the thick curve $\hat{U}_2$ vanishes.

\subsection*{Fig. 7}
Plot of $-\hat{U}_2$ for $\alpha=\pi/2$ (long dashes), and $\hat{U}_2$ for $\alpha=0$ (solid curve) and $\alpha=\pi/4$ (short dashes).

\subsection*{Fig. 8}
Plot of the reduced rotational friction coefficients $f_{Ra}/f_R$ (lower curve) and $f_{Rb}/f_R$ (upper curve) as functions of the aspect ratio $b/a$.

\subsection*{Fig. 9}
Plot of the reduced rotational swimming velocity $\hat{\Omega}_2$, defined in Eq. (4.14), as a function of $\xi=-\log_{10}\zeta$.

\newpage
\setlength{\unitlength}{1cm}
\begin{figure}
 \includegraphics{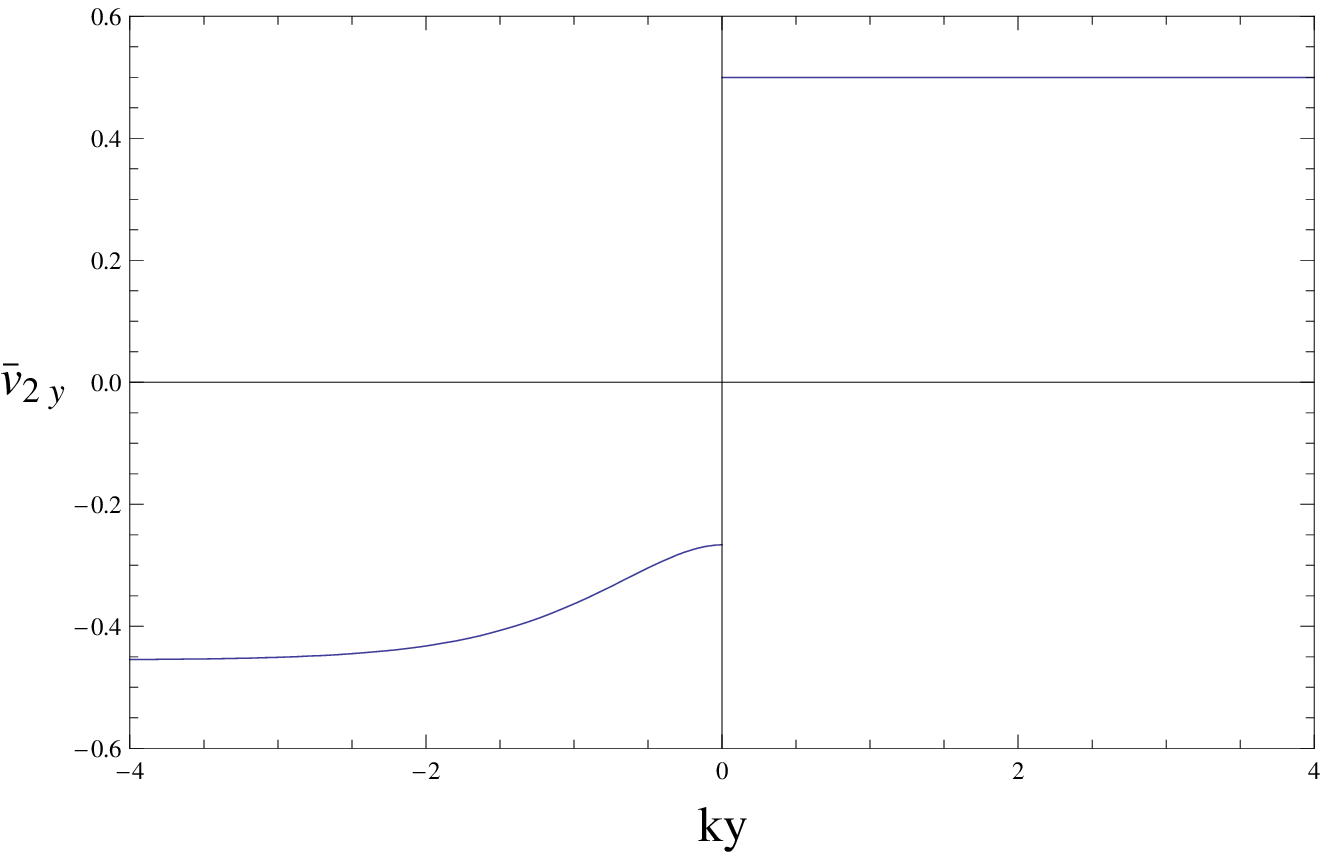}
   \put(-9.1,3.1){}
\put(-1.2,-.2){}
  \caption{}
\end{figure}
\newpage
\clearpage
\newpage
\setlength{\unitlength}{1cm}
\begin{figure}
 \includegraphics{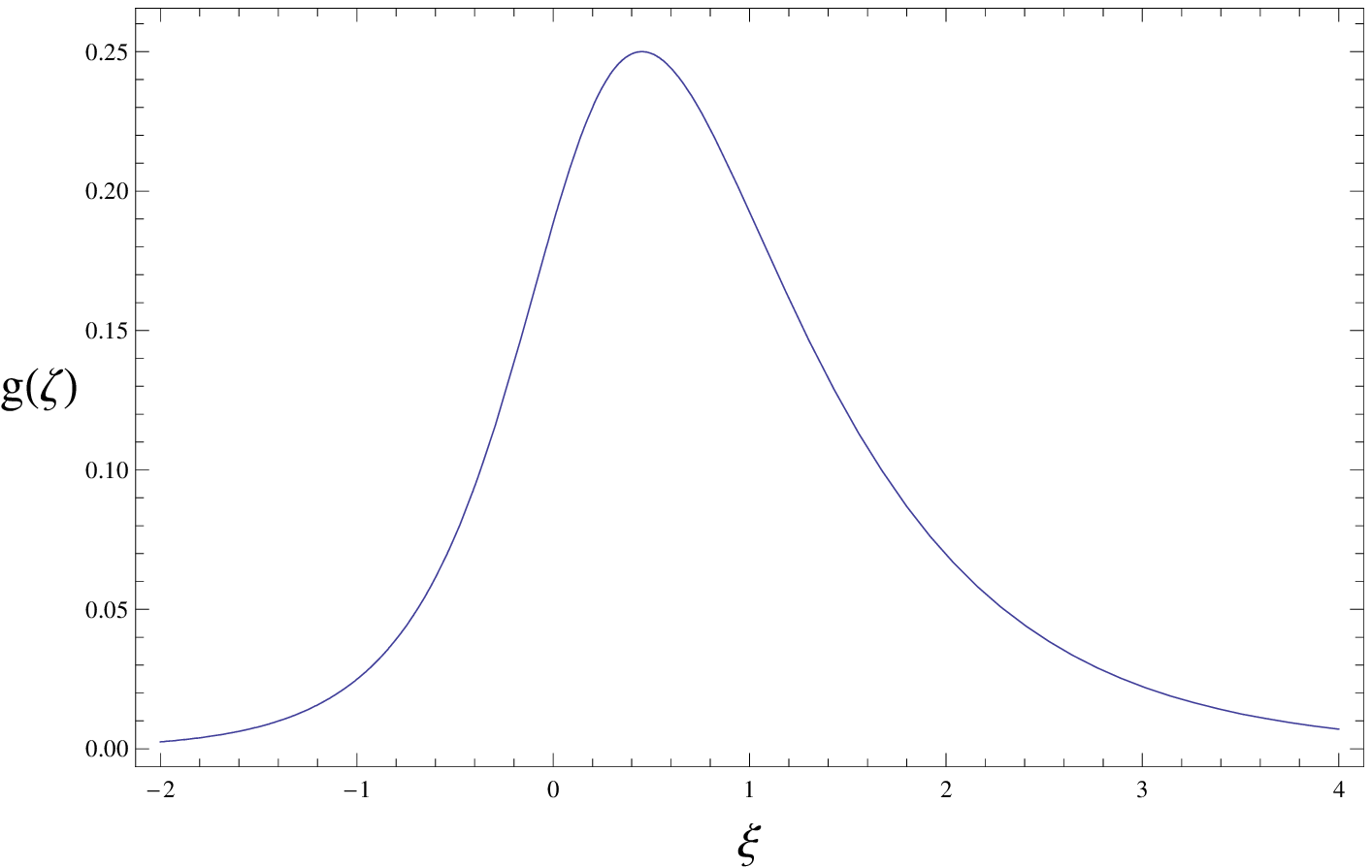}
   \put(-9.1,3.1){}
\put(-1.2,-.2){}
  \caption{}
\end{figure}
\newpage
\clearpage
\newpage
\setlength{\unitlength}{1cm}
\begin{figure}
 \includegraphics{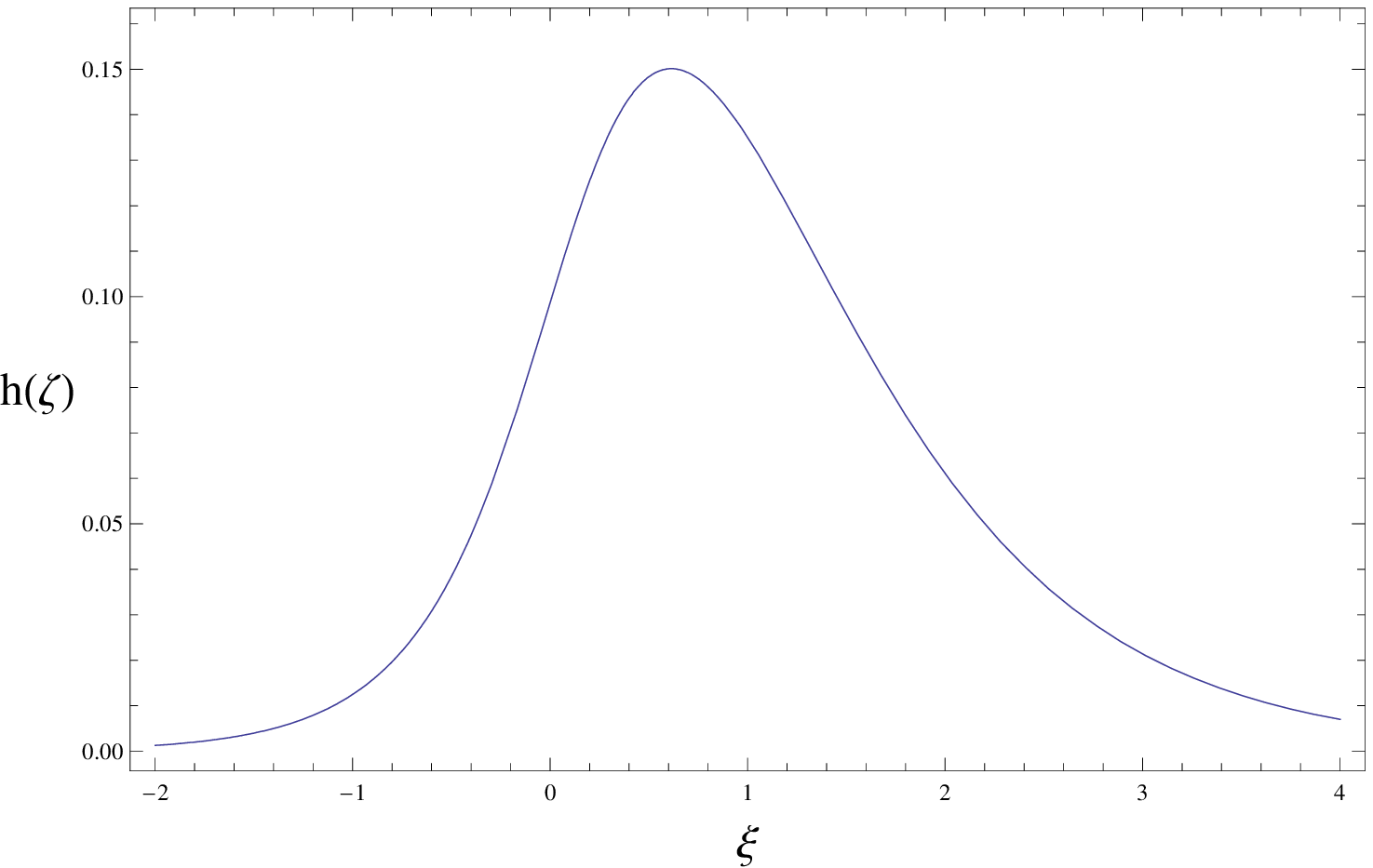}
   \put(-9.1,3.1){}
\put(-1.2,-.2){}
  \caption{}
\end{figure}
\newpage
\clearpage
\newpage
\setlength{\unitlength}{1cm}
\begin{figure}
 \includegraphics{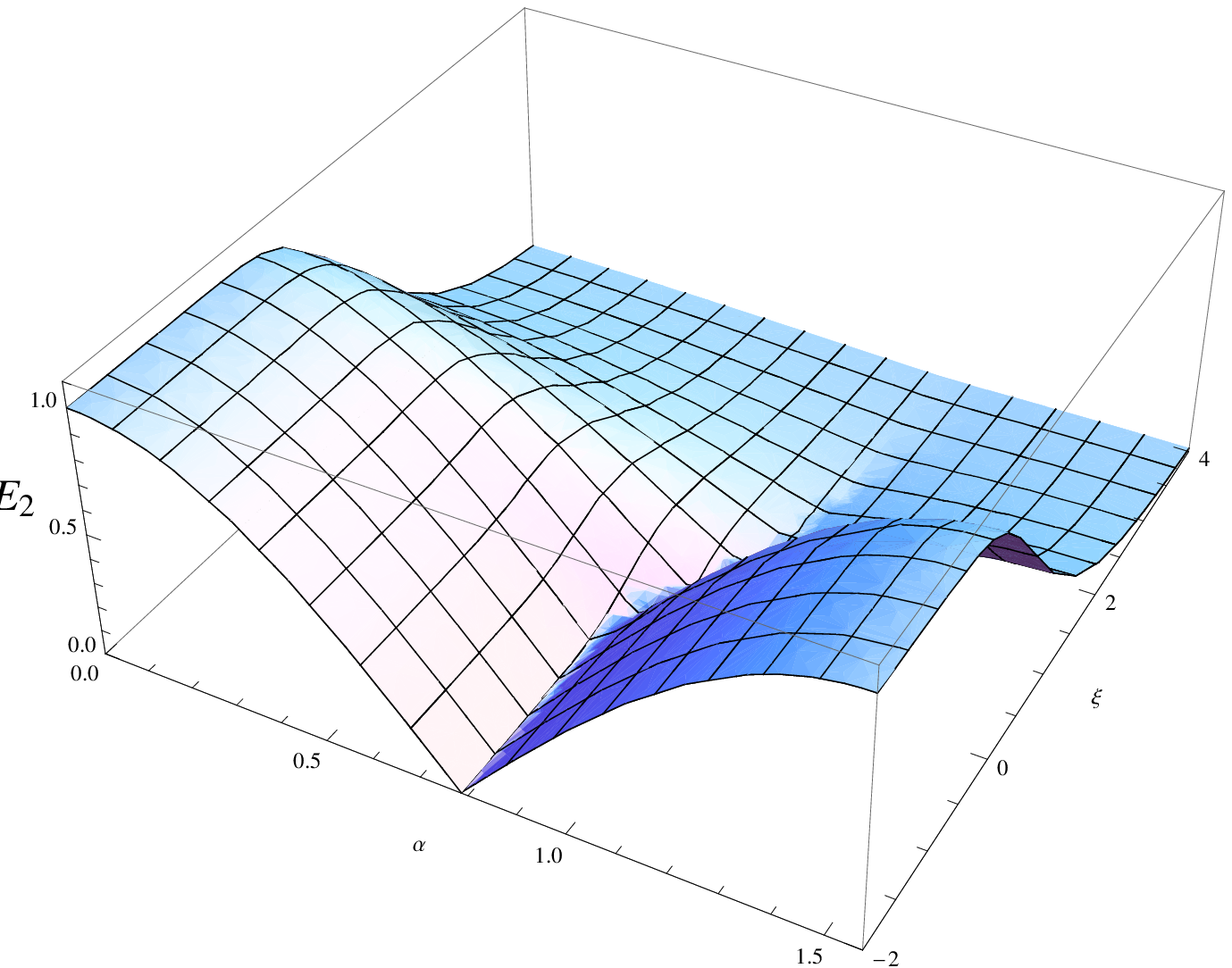}
   \put(-9.1,3.1){}
\put(-1.2,-.2){}
  \caption{}
\end{figure}
\newpage
\clearpage
\newpage
\setlength{\unitlength}{1cm}
\begin{figure}
 \includegraphics{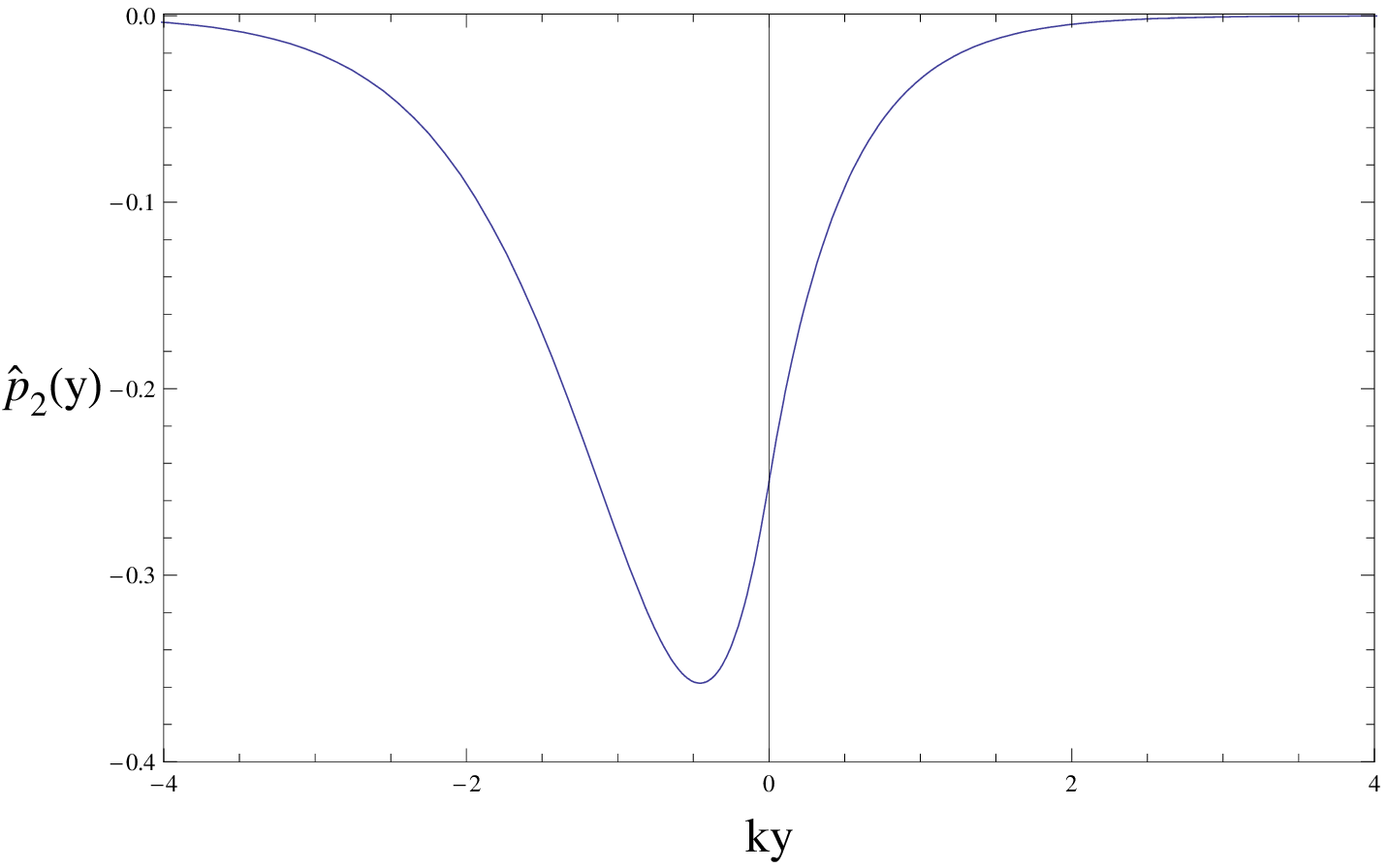}
   \put(-9.1,3.1){}
\put(-1.2,-.2){}
  \caption{}
\end{figure}
\newpage
\clearpage
\newpage
\setlength{\unitlength}{1cm}
\begin{figure}
 \includegraphics{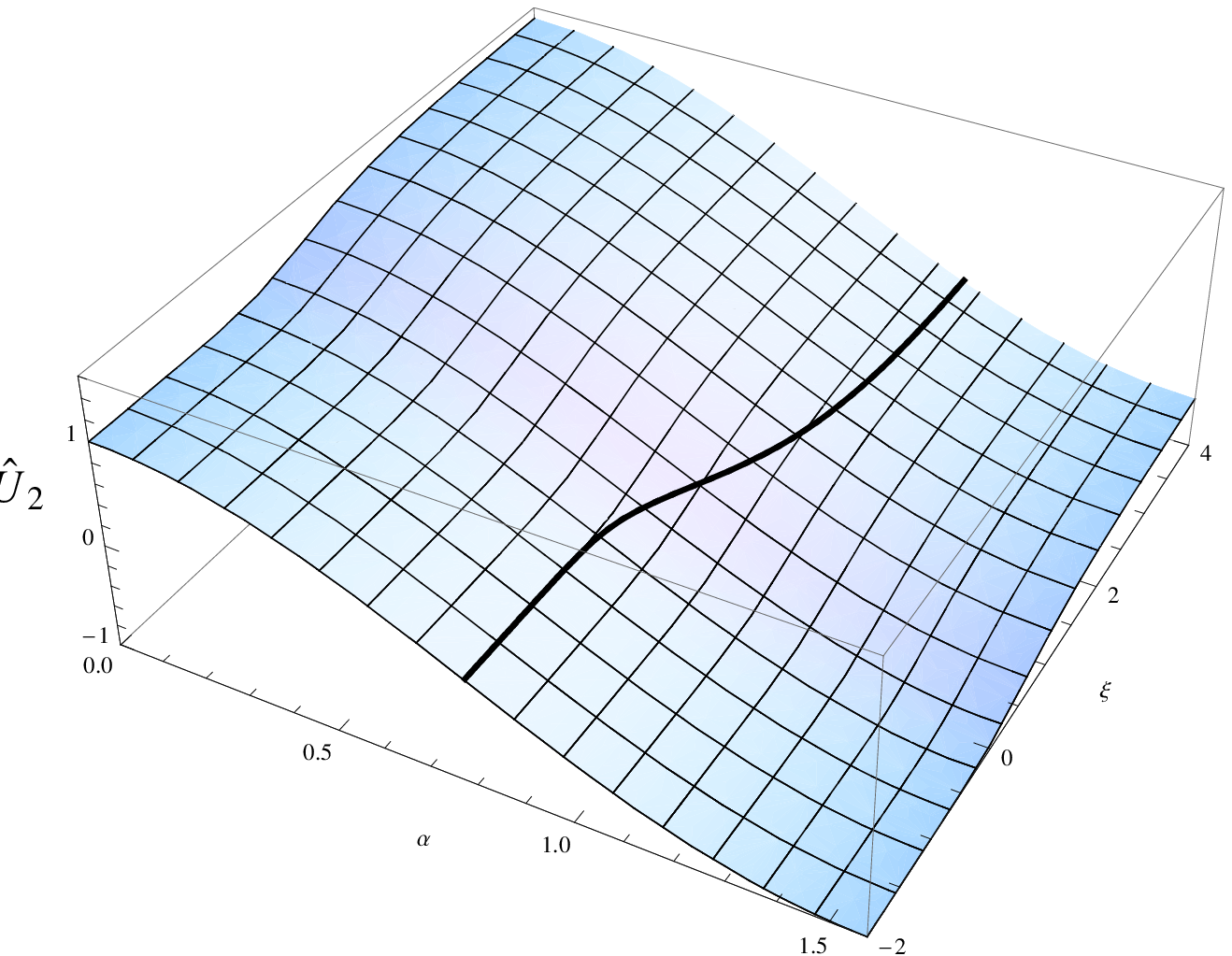}
   \put(-9.1,3.1){}
\put(-1.2,-.2){}
  \caption{}
\end{figure}
\newpage
\clearpage
\newpage
\setlength{\unitlength}{1cm}
\begin{figure}
 \includegraphics{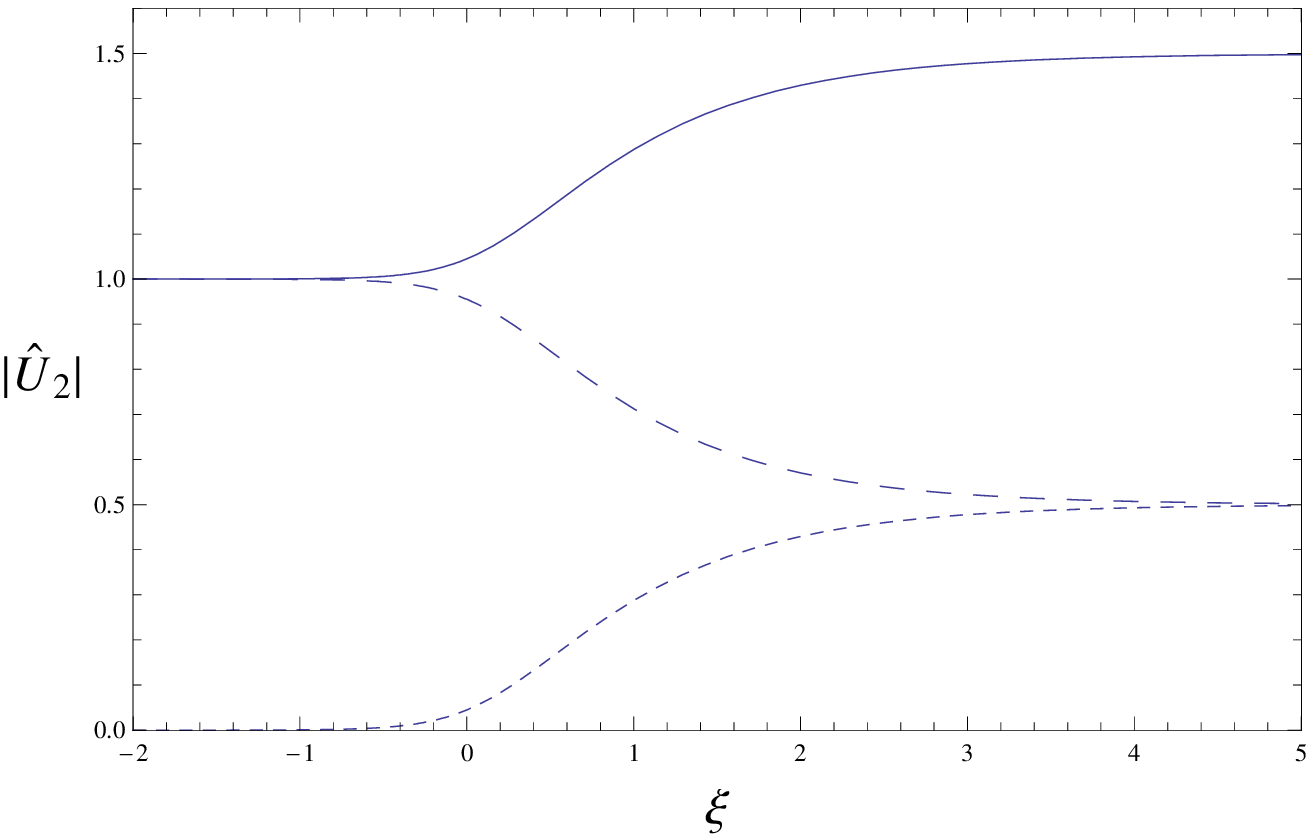}
   \put(-9.1,3.1){}
\put(-1.2,-.2){}
  \caption{}
\end{figure}
\newpage
\clearpage
\newpage
\setlength{\unitlength}{1cm}
\begin{figure}
 \includegraphics{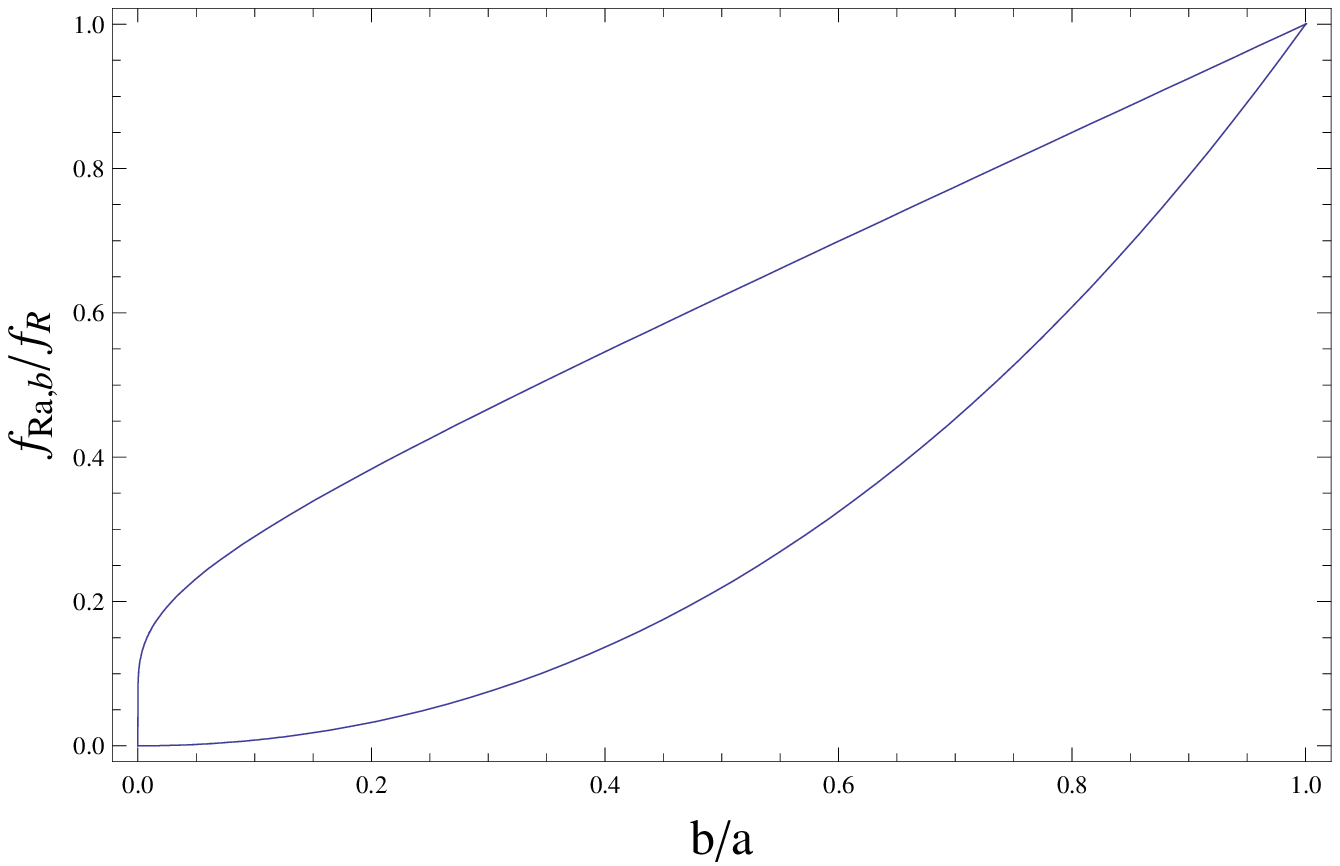}
   \put(-9.1,3.1){}
\put(-1.2,-.2){}
  \caption{}
\end{figure}
\newpage
\clearpage
\newpage
\setlength{\unitlength}{1cm}
\begin{figure}
 \includegraphics{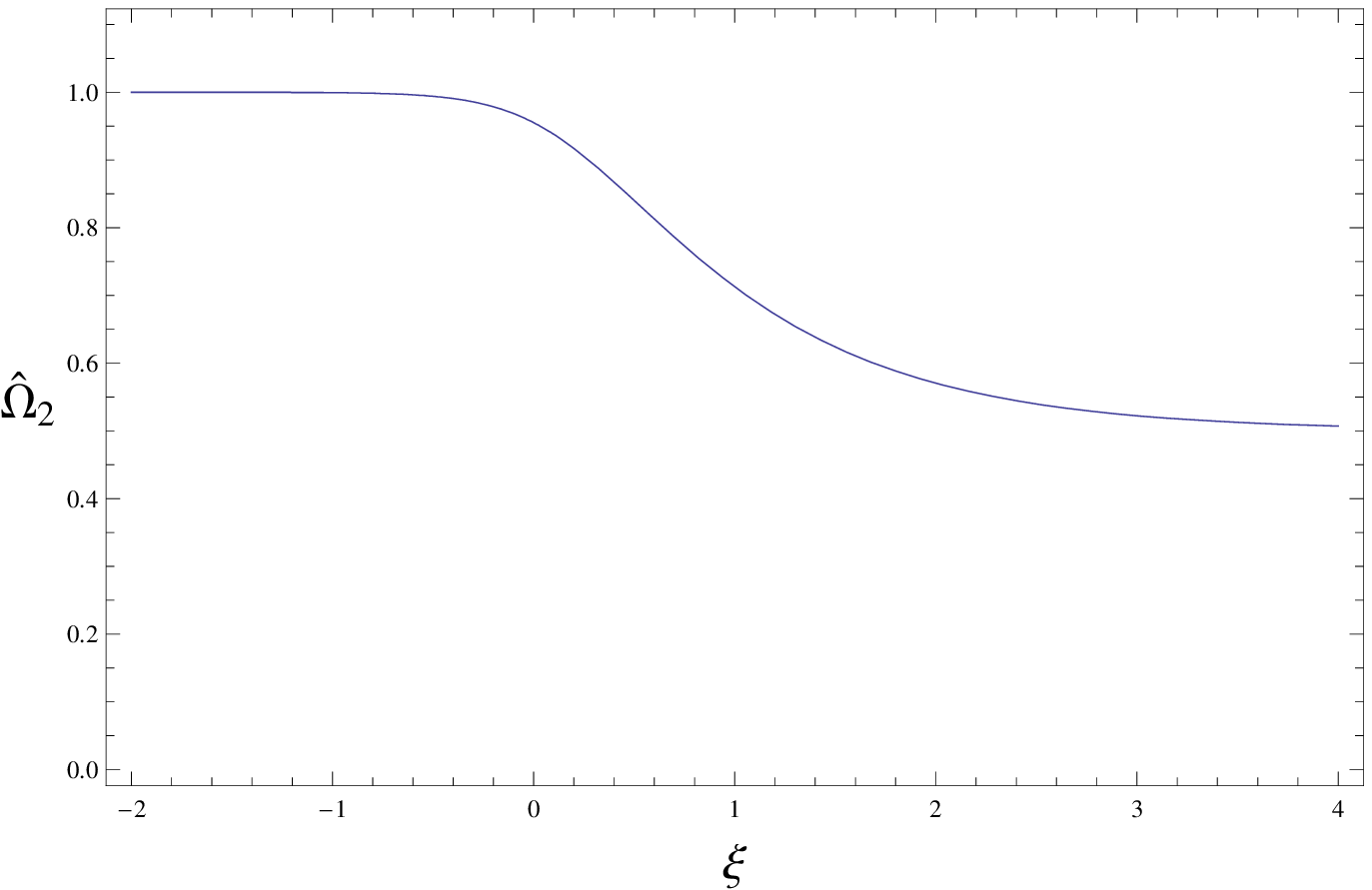}
   \put(-9.1,3.1){}
\put(-1.2,-.2){}
  \caption{}
\end{figure}
\newpage

\end{document}